\documentclass[aps,pra,twocolumn,groupedaddress,10pt]{revtex4}
\usepackage{amsmath}
\usepackage{amssymb}
\usepackage{graphicx}

\def\be{\begin{equation}} \def\ee{\end{equation}}
\def\bea{\begin{eqnarray}} \def\eea{\end{eqnarray}}

\def\nn{\nonumber}

\begin{document}
\title{A Quantized Inter-level Character in Quantum Systems}
\begin{abstract}
For a quantum system subject to external parameters, the Berry phase
is an intra-level property, which is gauge invariant module $2\pi$
for a closed loop in the parameter space and generally is non-quantized.
In contrast, we define a inter-band character $\Theta$ for a
closed loop, which is gauge invariant and quantized as integer
values.
It is a quantum mechanical analogy of the Euler character based
on the Gauss-Bonnet theorem for a manifold with a boundary.
The role of the Gaussian curvature is mimicked by
the difference between the Berry curvatures of the two levels, and
the counterpart of the geodesic curvature is the quantum geometric
potential which was proposed to improve the quantum adiabatic condition.
This quantized inter-band character is also generalized to
quantum degenerate systems.
\end{abstract}

\author{Chao Xu}
\author{Jianda Wu}
\author{Congjun Wu}
\affiliation{Department of Physics, University of California, San Diego, California 92093, USA}
\maketitle

\paragraph*{Introduction.---}
The study on time-dependent systems has great facilitated the exploration
of novel physics \cite{farhi2001quantum,das2008colloquium,eckardt2017colloquium,
georgescu2014quantum,kral2007colloquium,xiao2010berry,zhou2016accelerated,
bason2012high,hollenberg2012quantum,barends2015digital,ashhab2006decoherence}.
In particular, the research of the quantum adiabatic evolution
has led to a variety of important results, such as the quantum
adiabatic theorem \cite{Born1928,Schwinger1937,Kato1950},
the Landau-Zener transition \cite{Landau1932,Zener1932},
the Gell-Mann-Low theorem \cite{Gellmann1951},
and the Berry phase and holonomy \cite{berry1984proc,simon1983holonomy}.
It gives rise to many applications in quantum control and quantum computation \cite{Oreg1984,Schiemann1993,Pillet1993,Jones2000,Childs2001,Zheng2005,Lidar2009,Lidar2016,Santos2017}.
Another noteworthy example is the Berry phase and the corresponding gauge structure, which have been applied to condensed matter physics on revealing
novel phenomena, including the quantized charge pumping \cite{niu1990towards,PhysRevB.27.6083},
quantum spin Hall effect \cite{murakami2003,murakami20042,guo2008intrinsic},
quantum anomalous Hall effect \cite{PhysRevLett.61.2015}, and
electric polarization \cite{king1993theory,RevModPhys.66.899}.

The Berry phase equals the surface integral of the Berry curvature
over an area enclosed by a loop in the parameter space, while the
first Chern number corresponds to integrating the Berry curvature
over a closed surface.
According to the generalized Gauss-Bonnet theorem, the Chern number
is quantized.
The Chern number is very helpful in characterizing the topological
phase different from the ordinary ``phase" associated to the
symmetry breaking of local order parameters.
For example, the first Chern number characterizes
the quantization of Hall conductance \cite{thouless1982quantized,kohmoto1985topological}.
The Berry phase also has a deep relation to
the gauge field and differential geometry,
where it is viewed as a holonomy of the Hermitian line bundle \cite{simon1983holonomy}.
It can also be calculated by a line integral over a loop.
The integration result is independent of the linear velocity on the loop,
implying the geometric property of the Berry phase.
Wilczek and Zee further introduced the non-Abelian Berry phase \cite{wilczek1984appearance}, a generalization of the original
Abelian one \cite{berry1984proc}.
The non-Abelian Berry phase is presented
in the quantum degenerate system with a $U(N)$ gauge field,
which also has a deep relation to the topology,
such as the Wilson loop \cite{PhysRevD.10.2445} and the second Chern number.

The Berry phase is a consequence of the projection of the Hilbert
space to a particular level.
Around a closed loop, its value actually is gauge dependent but
remains invariant module $2\pi$.
On the other hand, the inter-level connection, i.e.,
the projection of the time-derivative of the state-vector of
one level to that of another one, is not well-studied.
An interesting application is the quantum geometric potential,
which has been applied to modify the quantum adiabatic
condition (QAC) \cite{PhysRevA.77.062114}, and its effect on quantum
adiabatic evolution has been experimentally detected \cite{Du2008}.

In this article, we construct a gauge invariant inter-level character
$\Theta$ based on the quantum geometric potential.
It is quantized in terms of integers, which can be
viewed as a counterpart of the Euler characteristic number for a
manifold with boundary.
The Gauss-Bonnet theorem says that there are two contributions to
the Euler characteristic numbers including the surface integral
of the Gaussian curvature and the loop integral of the geodesic
curvature along the boundary.
The quantum geometric potential plays the role of
the geodesic curvature, and the Berry curvature difference
between two levels is the analogy to the Gaussian curvature.
We also generalized the quantum geometric potential to the
case of degenerate quantum systems, and the quantized
character $\Theta$ can be constructed accordingly.

\paragraph*{Gauge invariant in non-degenerate quantum systems ---}
For non-degenerate quantum systems, an inter-level gauge
invariant, referred as
``quantum geometric potential'',
was introduced in literature \cite{PhysRevA.77.062114}.
Without loss of generality, we start with a non-degenerate $N$-level
Hamiltonian $\hat{H}(\vec{\lambda}(t))$ controlled by a real $l$-vector
$\vec{\lambda}(t) = \{\lambda_1(t), \lambda_2(t),\cdots,\lambda_l(t) \}$
as a function of time $t$.
At each fixed $t$, a set of orthonormal eigenfunctions
$|\phi_m(\vec{\lambda})\rangle$ associated with the
eigenvalues $E_m(\vec{\lambda})$ are determined by
$\hat{H}(\vec{\lambda})|\phi_m(\vec{\lambda})\rangle =
E_m(\vec{\lambda})|\phi_m(\vec{\lambda})\rangle$, ($m=1,2,\cdots,N$).
The Berry connection for each energy level is defined as
$\mathcal{A}_{m}^{\mu}=i\langle\phi_m(\vec{\lambda})
|\partial_{\lambda_\mu}|\phi_m(\vec{\lambda})\rangle$ ($\mu = 1, 2, \cdots, l$).
Consequently, quantum geometric potential arises as,
\be
\Delta_{{\rm{ND}},mn} = \mathcal{A}_{n} - \mathcal{A}_{m} + \frac{d}{dt}\arg \langle\phi_m|\dot{\phi}_n\rangle,
\label{eq:ndqgp}
\ee
where ND denotes the non-degenerate systems, and the ``$\cdot$"
illustrates the time-derivative.
In addition, $\mathcal{A}_{m} \equiv \mathcal{A}_{m}^{\mu} \dot{\lambda}_{\mu} $
(in this paper the repeated indices imply the summation).
The adiabatic solution to the time-dependent Schr\"odinger equation,
$i \partial_t |\eta_{m}^a(\vec{\lambda}(t))\rangle
= \hat{H}(\vec{\lambda}(t))|\eta_{m}^a(\vec{\lambda}(t))\rangle$ is
\begin{equation}\label{eta}
|\eta_{m}^a(t)\rangle = \exp\{-i\int_0^t E_{m}(\tau)d\tau\}|\tilde{\phi}_{m}^a(t)\rangle,
\end{equation}
with
$|\tilde{\phi}_{m}^a(t)\rangle = \exp\{\int i\mathcal{A}_m dt \}|\phi_{m}^a(t)\rangle$,
if the initial state $|\eta_{m}^a(0)\rangle = |\phi_{m}^a(0)\rangle$.
Then $\Delta_{{\rm{ND}},mn}$ can be also defined as
\be\label{tild}
\Delta_{{\rm{ND}},mn} = \frac{d}{dt}\arg \langle\tilde{\phi}_m|\dot{\tilde{\phi}}_n\rangle.
\ee

$\Delta_{{\rm ND},mn} $ is gauge invariant under an arbitrary
local $U(1)\otimes U(1)$ gauge transform with
$|\phi_{m(n)}(t)\rangle\rightarrow e^{i\alpha_{m(n)}(t)}|\phi_{m(n)}(t)\rangle$
where $\alpha_{m(n)}(t)$ are smooth scalar functions.
In the spin-$\frac{1}{2}$ system coupled to an external time-dependent
magnetic field, $\Delta_{\rm{ND}}$ is equivalent to the
geodesic curvature of the path of the magnetic field orientation
on the Bloch sphere, implying its geometric implications.
When applying $\Delta_{{\rm{ND}},mn}$ to the time-dependent system,
an improved QAC for the non-degenerate system can be established
for $n\neq m$ \cite{PhysRevA.77.062114},
\be
\frac{{\left| {\langle\phi_m|\dot{\phi}_n\rangle } \right|}}
{{\left| {E_m (t) - E_n (t) + \Delta _{{\rm{ND}},mn} (t)} \right|}}
\ll 1 \label{eq:U1adia},
\ee
which indicates ${E_m (t) - E_n (t) + \Delta_{{\rm{ND}},mn} (t)} $
is more appropriate to describe the instantaneous energy gaps.

\paragraph*{A quantized character in non-degenerate system---}
We introduce a new quantized gauge invariant character
$\Theta$ based on the quantum geometric potential as an analogy to Gauss-Bonnet
theorem with boundary.
\begin{figure}
\includegraphics[scale=0.35]{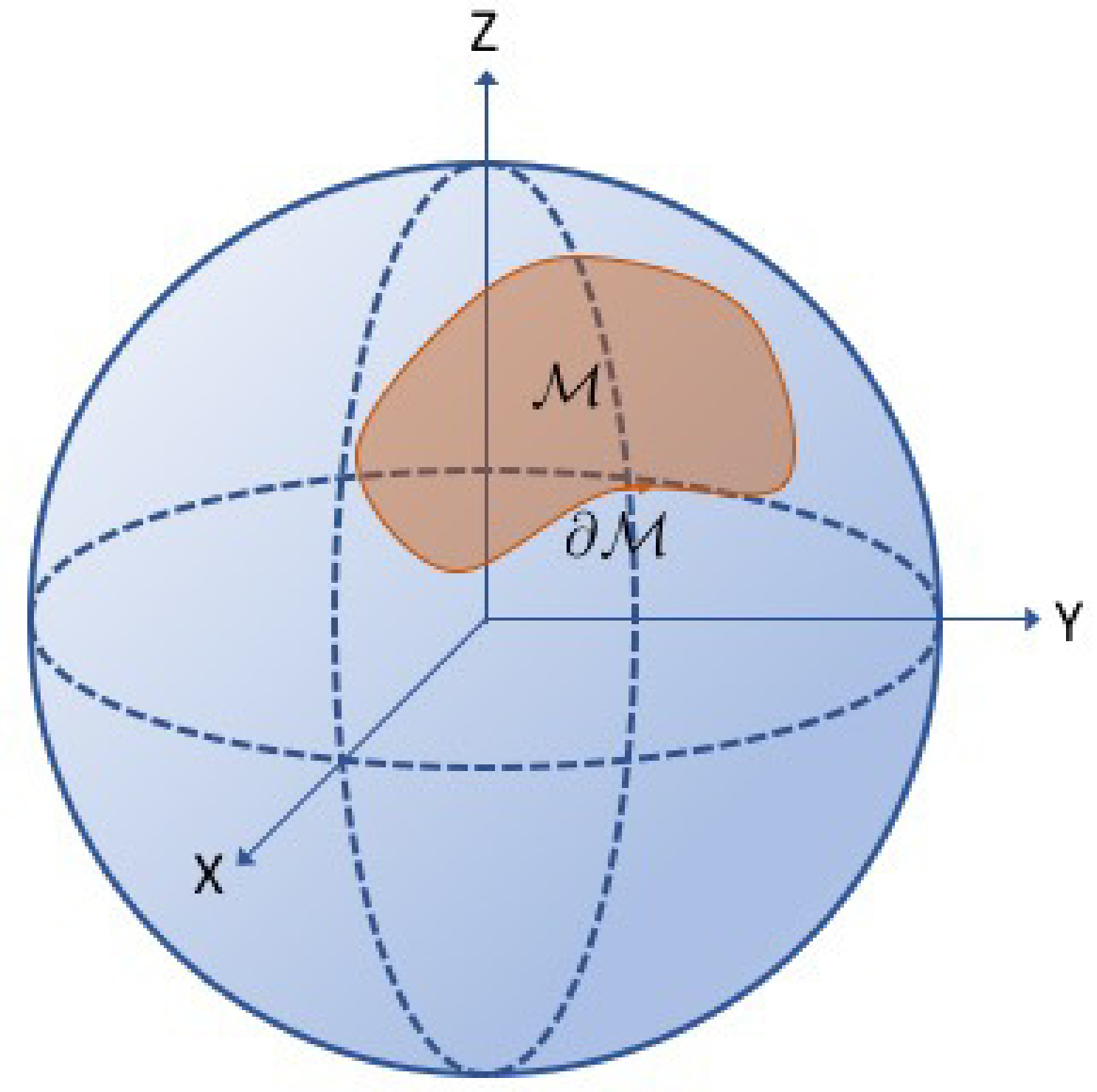}
\caption{The region $\mathcal{M}$ on the $\mathbb{S}^2 $
Bloch sphere with a smooth boundary $\partial\mathcal{M}$.}
\label{f1}
\end{figure}
For simplicity, we begin with a two-level system controlled
by a real 3-vector $\vec{\lambda}(t)$.
At each time $t$, there exist a pair of eigenfunctions
$|\phi_{\pm}(\vec{\lambda}(t))\rangle$ associated with
the eigenvalues $E_{\pm}(\vec{\lambda}(t))$.
Define $\omega = (\mathcal{A}_{-}^{\mu}-\mathcal{A}_{+}^{\mu})
d\lambda^{\mu}$, and $\mathcal{F} = d\omega$ with $d$ being
the exterior derivative.
Explicitly, $\mathcal{F}$ is carried out as
$\mathcal{F} = \mathcal{F}_- -\mathcal{F}_+ $,
where $\mathcal{F}_{\pm} = \frac{1}{2}F_{\pm}^{\mu\nu} d\lambda^{\mu}\wedge d\lambda^{\nu}$
with $F^{\mu\nu}_{\pm} = \partial^\mu
\mathcal{A}^{\nu}_{\pm} - \partial^\nu \mathcal{A}^{\mu}_{\pm}$.
A novel quantized character $\Theta$ is defined as
\begin{eqnarray}\label{theta}
2\pi\Theta &=&\int_{\mathcal{M}}\mathcal{F} -
\int_{\partial\mathcal{M}} \Delta dt\nn \\
&=&\Phi_+ - \Phi_-
-\int_{\partial \mathcal{M}}d ~{\rm{arg}}\langle\phi_+|\dot{\phi}_-\rangle,
\end{eqnarray}
where $\Delta$ is the gauge invariant
in Eq.~(\ref{eq:ndqgp}) for the non-degenerate systems, and
$\Phi_\pm=
\int_{\partial \mathcal{M}}\mathcal{A}_{\pm}^{\mu}d\lambda_{\mu} - \int_{\mathcal{M}}\mathcal{F}_{\pm}$.
Since $\cal F$ and $\Delta$ are both locally gauge invariant,
$\Theta$ is also gauge invariant.

\begin{figure}
\includegraphics[scale=0.43]{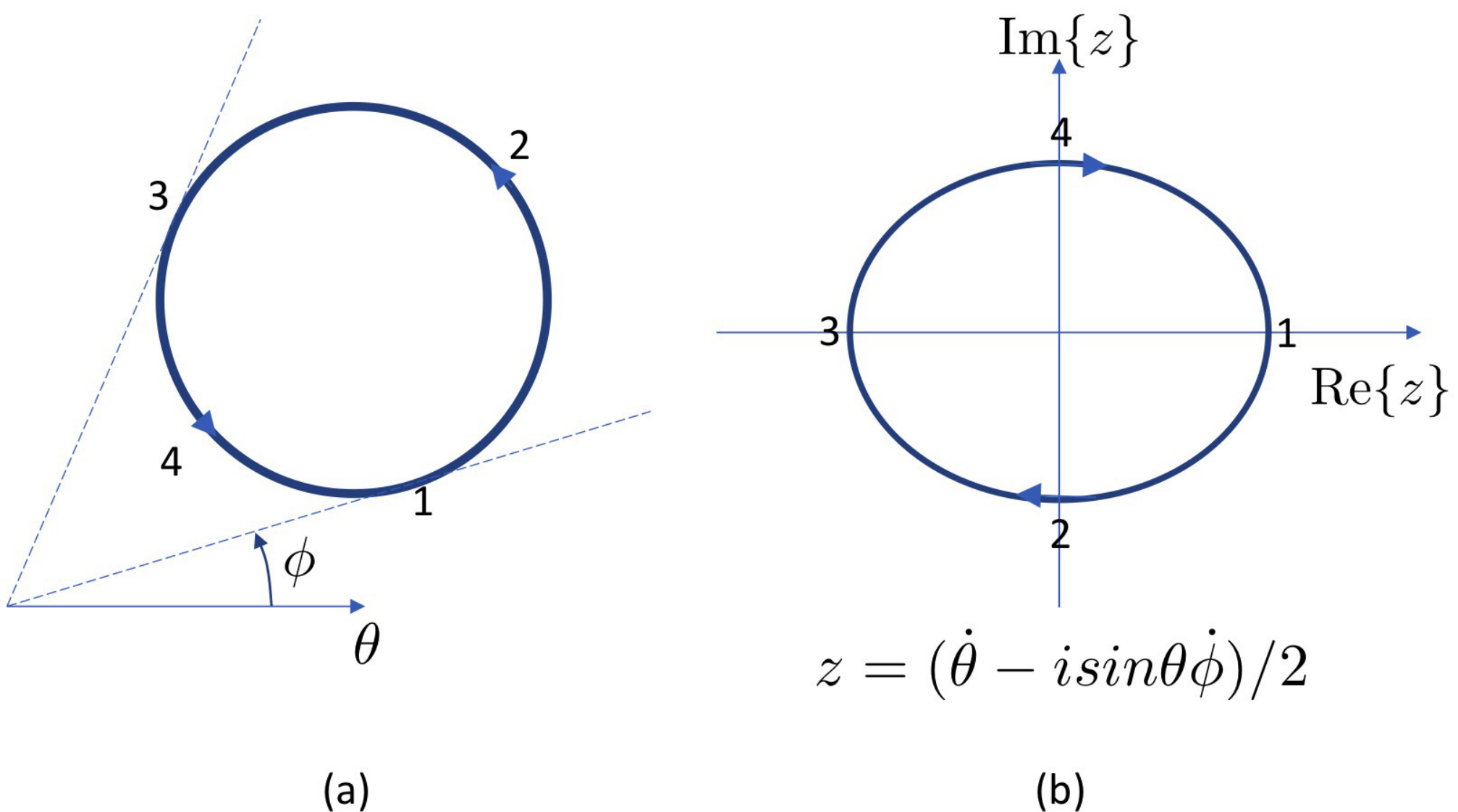}
\caption{(a) The top view of a closed curve on the Bloch sphere
in the vicinity of the north pole.
$\theta$ and $\phi$ represent the radial and angular coordinates,
respectively.
(b) The corresponding curve $z(t)$ in the complex plane with
$z(t)=\langle\phi_+|\dot{\phi}_-\rangle = (\dot{\theta}-i\sin\theta\dot{\phi})/2$.}
\label{rz}
\end{figure}

To show the quantization of $\Theta$, we first consider a
simple example of a two-level problem
with the Hamiltonian $\hat{H}(t) = B\hat{n}(t)\cdot \vec{\sigma}$.
Here $\hat{n}$ is a 3D unit vector, and the whole parameter space
is the Bloch sphere.
If $\hat{n}(t)$ concludes a region $\mathcal{M}$
on the Bloch sphere with a smooth boundary $\partial\mathcal{M}$ (Fig.~\ref{f1}), then $\Theta$ is quantized.
Consider the transition term $\langle \phi_{+}|\dot{\phi}_-\rangle$
from the ground state to the excited state, which is a complex number.
The corresponding $\mathcal{F}$
is the Berry curvature difference between the ground and excited states.
To explicitly calculate $\Theta$, we can work in a give gauge that
$|\phi_-(\theta,\phi)\rangle = (\sin\frac{\theta}{2}e^{-i\phi},-\cos\frac{\theta}{2})^T$ and
$|\phi_+(\theta,\phi)\rangle = (\cos\frac{\theta}{2}e^{-i\phi},\sin\frac{\theta}{2})^T$.
Under this gauge, $\Phi_+=2\pi$ if $\partial \mathcal{M}$
encloses the north pole, and $\Phi_-=-2\pi$ if it encloses
the south pole.
Otherwise $\Phi_\pm=0$.
Meanwhile ${\rm{arg}}\langle\phi_+|\dot{\phi}_-\rangle
= {\rm{arg}}\left((\dot{\theta}-i\sin\theta\dot{\phi})/2\right)$.
When $\vec{\lambda}(t)$ completes a close loop $\partial \mathcal{M}$,
correspondingly, $z(t)=\langle\phi_+|\dot{\phi}_-\rangle$
defines a close curve in complex plane.
The winding number of $z(t)$ relative to the origin
is defined as $W[z] = \int_{\partial \mathcal{M}}d ~{\rm{arg}}\langle\phi_+|\dot{\phi}_-\rangle$
as shown in Fig.~\ref{rz}.
If $\partial \mathcal{M}$ does not enclose the north or south pole,
$\Phi_\pm$ do not contribute, and
$W[\langle\phi_+|\dot{\phi}_-\rangle]$ contributes $-2\pi$,
such that $\Theta = 1$.
After a similar analysis for other situations, one can conclude
that $\Theta=1$ for any region $\mathcal{M}$ on the sphere.

\begin{figure}
\includegraphics[scale=0.45]{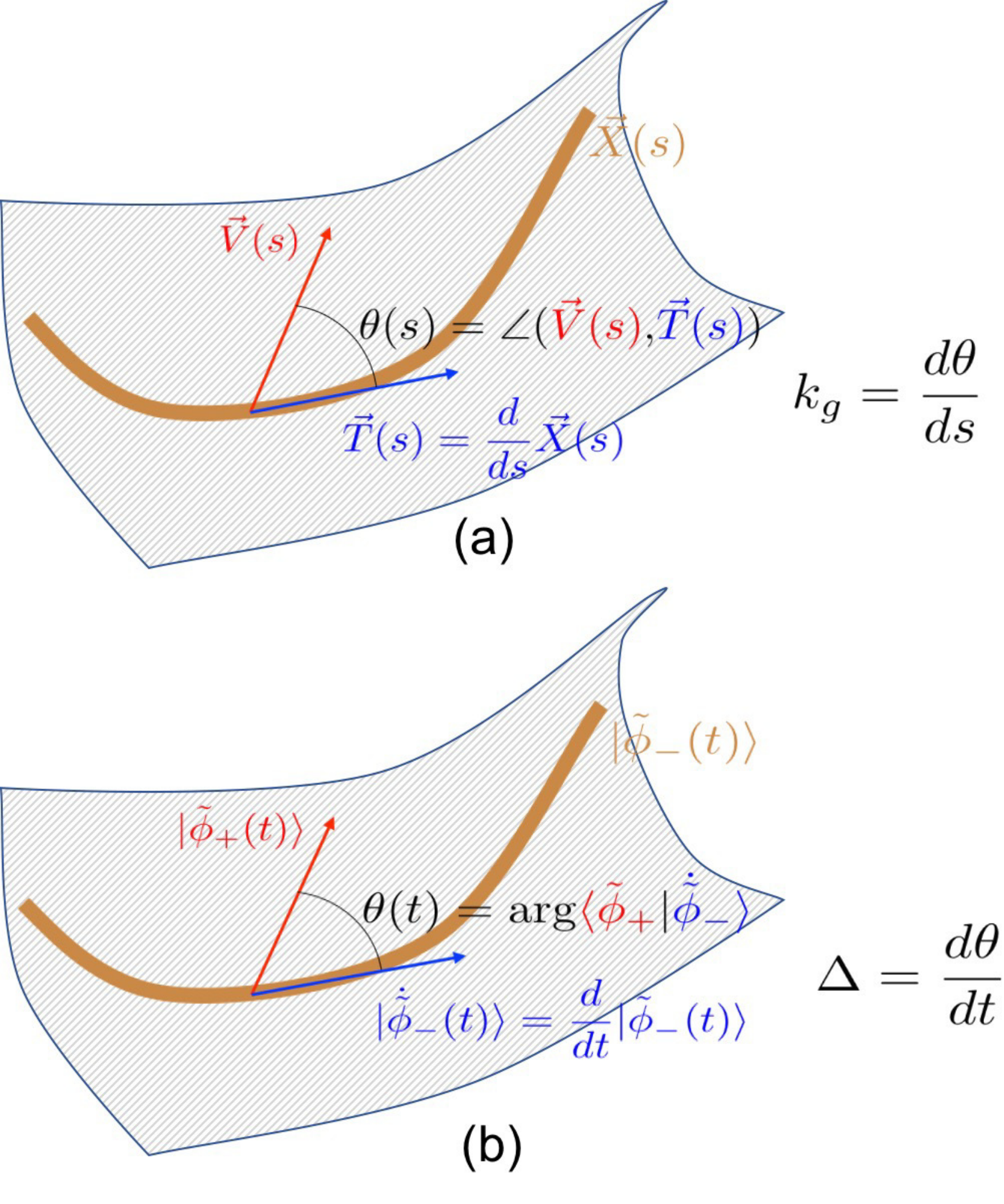}
\caption{(a) A curve $\vec{X} (s)$ is plotted on a 2D manifold (shaded area) in the
3D real space.
$\vec{V}(s)$ lives in the tangent space, and is parallelly transported
along the curve.
$\vec{T}(s) = \frac{d}{ds}\vec{X}(s)$ is the velocity vector, and
$\theta$ is the angle between $\vec{V}$ and $\vec{T}$.
The geodesic curvature $k_g=d\theta/ds$.
(b) The trajectory of $|\tilde{\phi}_-(t)\rangle$ is sketched
in the Hilbert space.
$|\tilde{\phi}_+(t)\rangle$	is a parallelly transported ``tangent"
vector along the ``curve".
$|\dot{\tilde{\phi}}_-(t)\rangle$ is the velocity vector, which is the derivative of of the ``curve".
The gauge invariant $\Delta=d\theta/dt$, where $\theta = \arg\langle \tilde{\phi}_+|\dot{\tilde{\phi}}_-\rangle$.}
\label{f2}
\end{figure}

For a general non-degenerate model, we can define the quantized
character $\Theta$ between any two different energy levels
$E_\pm$ associated with a closed curve in the parameter space.
According to the Stokes theorem, $\Phi_\pm$ count the singularities
of Berry connections $\cal A_\pm^\mu$ in the region $\mathcal{M}$,
e.g. the number of the Dirac strings, hence, they are quantized.
The winding number of $z(t)$ relative to the origin is also quantized.
Therefore, $\Theta$ is quantized for any situation.

Below we demonstrate the similarities between the quantized
character $\Theta$ and the Euler number in the Gauss-Bonnet theorem.
For a 2D compact Riemannian manifold $\mathcal{M}$ with a smooth
boundary $\partial \mathcal{M}$, the Gauss-Bonnet theorem reads
\begin{equation}\label{GB}
\int_{\mathcal{M}} G dA + \int_{\partial \mathcal{M}} k_g ds = 2\pi \chi(\mathcal{M}),
\end{equation}
where $G$, $k_g$ and $\chi(\mathcal{M})$ are the Gaussian curvature,
geodesic curvature of $\partial \mathcal{M}$, and the Euler number
of $\mathcal{M}$, respectively.
For quantum systems (e.g. a spin-1/2 problem in an external magnetic
field), each point in the parameter space has an associated Hilbert
space, i.e., the bundle.
The Gauss-Bonnet theorem is generalized to characterize the bundle
by the Chern number.

The gauge invariant $\Delta_{\rm{ND}}$ defined in Eq.~(\ref{eq:ndqgp}) is the analogy
to the geodesic curvature $k_g$ in Eq.~(\ref{GB}).
To explain this, we plot a curve $\vec{X}(s)$ on a 2D manifold in $\mathbb{R}^3$ as shown in the Fig.~\ref{f2}a, which is parameterized by the arc length $s$.
$\vec{X}(s)$ represents the displacement vector for a point on the curve, then $k_g$ is a geometric quantity depending on both the manifold and the curve.
The geodesic curvature $k_g$ reflects the deviation of the curve from
the local geodesics.
Choose a vector function $\vec V(s)$ living in the tangent space at the position $\vec X(s)$ and is parallelly transported along the curve.
Then $k_g = d\theta/ds$, where $\theta$ is the angle between the velocity vector $\vec{T}=d\vec{X}/ds$ and $\vec V(s)$.

The similarity between $\Delta_{{\rm{ND}}}$ and $k_g$ is illustrated
in Fig. \ref{f2} b.
Following Eq. \ref{tild}, the trajectory of $|\tilde{\phi}_-(t)\rangle$
is viewed as a curve in the Hilbert space.
Due to the Berry phase, $|\dot{\tilde{\phi}}_-(t)\rangle$ is actually orthogonal to $|\tilde{\phi}_-(t)\rangle$.
For the two-level system $|\tilde{\phi}_+\rangle$ is just $|\dot{\tilde{\phi}}_-(t)\rangle$ up to a complex factor,
hence, $|\tilde{\phi}_+(t)\rangle$ is the analogy to the parallelly
transported ``tangent" vector field along the ``curve".
Consequently, the gauge invariant term $\Delta_{\rm{ND}} = d\theta/dt$ is
determined by the derivative of the angle
$\theta = \arg\langle \tilde{\phi}_+|\dot{\tilde{\phi}}_-\rangle$
over time.
Therefore, therefore Eq.~(\ref{theta}) can be viewed as a quantum analogy
to the Gauss-Bonnet theorem described in Eq.~(\ref{GB}).

There exist fundamental differences between the gauge invariant $\Delta_{\rm{ND}}$ and the usual Berry connection.
The integral of  $\Delta_{\rm{ND}}$ over a close loop is
gauge invariant and single-valued.
In contrast, the Berry connection is {\it not} gauge invariant locally,
and the Berry phase for a closed loop evolution is gauge invariant
but multiple valued module $2\pi$.
The Berry connection and the Berry phase are intra-subspace quantities associated with one energy level, while
$\Delta_{\rm{ND}}$ is an inter-subspace property associated with
two different energy levels.

\paragraph*{A quantized character in degenerate systems ---}
The gauge invariant quantized character $\Theta$ studied above
can also be extended to the degenerate systems.
For this purpose, the gauge invariant $\Delta_{\rm{ND}}$ is generalized
to the case with degeneracy, which is defined between
two eigenspaces associated
with two different degenerate energy levels.
We first consider a special case that a Hamiltonain $\hat{H}(\vec{\lambda})$
possessing $N$ energy levels $E_m(\vec{\lambda})~(m=1,2,\cdots, N)$,
each of which is $L$-fold degenerate.
The situation for energy levels possessing different degeneracies
is discussed in Appendix C.

For each energy level $m$, there is a set of instantaneous
orthonormal eigenstates $|\phi^a_m(\vec{\lambda})\rangle$,
satisfying $\hat{H}(\vec{\lambda})|\phi^a_m(\vec{\lambda})\rangle =
E_m(\vec{\lambda})|\phi^a_m(\vec{\lambda})\rangle$ ($a = 1, 2, \cdots, L$).
If the system evolves adiabatically starting from the initial
state
$|\eta_{m}^a(\vec{\lambda}(0))\rangle = |\phi_{m}^a(\vec{\lambda}(0))\rangle$,
then the adiabatic solution to the time-dependent Schr\"odinger equation,
$i \partial_t |\eta_{m}^a(\vec{\lambda}(t))\rangle
= \hat{H}(\vec{\lambda}(t))|\eta_{m}^a(\vec{\lambda}(t))\rangle$ is
\begin{equation}\label{eta}
|\eta_{m}^a(t)\rangle = \exp\{-i\int_0^t E_{m}(\tau)d\tau\}|\tilde{\phi}_{m}^a(t)\rangle
\end{equation}
with
$|\tilde{\phi}_{m}^a(t)\rangle = |\phi_{m}^b(t)\rangle[\Omega_{m}(t)]^{ba}$.
The non-Abelian Berry phases $\Omega_{m}$ and the correspoinding
Berry connections $\mathcal{A}_{m}^{\mu}$ are defined as
\bea
\Omega_{m}(t) &=& \mathcal{P}\left\{\exp
\Big\{i\int_{\vec{\lambda}(0)}^{\vec{\lambda}(t)} \mathcal{A}^\mu_{m}
d\lambda^{\mu}\Big\}\right \}, \\
\label{A}
\mathcal{A}_{m}^{\mu}(\vec{\lambda})^{ab} &=& i\langle\phi_{m}^{a}(\vec{\lambda})|\partial_{\lambda^{\mu}}|\phi_{m}^{b}(\vec{\lambda})\rangle
\label{omega},
\eea
where $\mathcal{P}$ means path-ordering \cite{wilczek1984appearance}.
The exact time-dependent solution can be expanded
as $|\psi(t)\rangle = c_{m}^a(t) |\eta_{m}^a(t)\rangle $, then one obtains
\begin{equation}
\dot{c}_{m}^{a}(t) = - \sum_{n\neq m}\exp
\Big \{i\int_0^t\epsilon_{mn}(\tau)d\tau\Big\}
~(\Omega_{m}^{\dag}T_{mn}\Omega_{n})^{ab}c_{n}^{b}(t),
\label{eq:time}
\end{equation}
where $\epsilon_{mn}(\tau) = E_m(\tau) - E_n(\tau)$
(details in the Appendix A).
The transition matrices $T_{mn}$ are followed by
\begin{equation}\label{tran}
	T^{ab}_{mn} = \langle\phi_{m}^a(\vec{\lambda})
|\partial_t|\phi_{n}^b(\vec{\lambda})\rangle,
\end{equation}
where $a$ and $b$ denote the row and column indices
of the matrix $T_{mn}$, respectively,
with $m$ and $n$ being energy level labels.

To figure out the gauge invariant in the degenerate case, we extract
the ``phase" from $\Omega_{m}^{\dag}T_{mn}\Omega_{n}$, i.e., the
counterpart of $\Delta_{\rm ND, mn}(t)$ in Eq.~(\ref{eq:ndqgp}).
The ``phase'' of $T$ is defined as $\theta_T= \frac{-i}{L}
{\rm{Tr}} [{\ln(UV^\dag)}]$, where $U$ and $V$ are unitary matrices
from $T's$ singular value decomposition,
$T_{mn} = U_{mn}S_{mn}V_{mn}^{\dag}$, and $S_{mn}$ is
a diagonal real matrix with non-negative elements.
We assume all the singular values of $T$ are positive
(The details are in the Appendix B).
The ``phase" of $\Omega_{m}$ is $\frac{1}{n} {\rm{Tr}}\{\int\mathcal{A}_{m}
d\tau\}$ where $\mathcal{A}_{m}= \mathcal{A}^\mu_{m} \dot{\lambda}^{\mu}$,
i.e., because $\Omega_{m}$ can be expressed as
$\exp \{\int \frac{i}{n} {\rm{Tr}}\{\mathcal{A}_{m}\}d\tau\}\bar{\Omega}_{m}$
where $\det\bar{\Omega}_{m}=1$.
Then the gauge invariant in the degenerate systems is defined as
\be
\Delta_{{\rm{D}},mn} = \frac{1}{L}  {\rm{Tr}}
\left\{\mathcal{A}_{n}-\mathcal{A}_{m}-i\frac{d}{dt}\ln(U_{mn}V_{mn}^\dag)
\right\},
\label{delta}
\ee
or, in a compact form
\begin{equation}
	\Delta_{{\rm{D}},mn} = -\frac{i}{L} {\rm{Tr}}\{\dot{X}_{mn}X_{mn}^{\dag}\}
\label{moregeneral}
\end{equation}
with $X_{mn}(\vec{\lambda}(t)) = \Omega_{m}^{\dag}U_{mn}V_{mn}^{\dag}\Omega_{n}$
(here ``D" denotes the degenerate systems).
The ``phase'' of $\Omega_m^{\dag} T_{mn} \Omega_n$ is defined as
$\int i\Delta_{{\rm{D}},mn} d\tau$, and Eq.~(\ref{eq:time})
can be rewritten as
\begin{eqnarray}\label{mean}
\dot{c}_{m}^{a} &=& - \sum_{n\neq m}
\exp \left \{i\int_0^t(\epsilon_{mn}(\tau)+\Delta_{\rm{D},mn}(\tau)) d\tau
\right \} \nn \\
&\times& (\bar{\Omega}_m^\dag \bar{U}_{mn}S_{mn}\bar{V}_{mn}^\dag
\bar{\Omega}_n)^{ab}c_{n}^{b}(t).
\end{eqnarray}
Similar as $\Delta_{{\rm{ND}},mn}$ in non-degenerate situations,
$\Delta_{{\rm{D}},mn}$ provides a proper correction for the
instantaneous energy gaps for the degenerate systems.
With the introduction of $\Delta_{{\rm{D}},mn}$, a modified QAC
is discussed in Appendix A.

$\Delta_{{\rm D},mn}$ is $U(L)\otimes U(L)$  gauge invariant under any
two independent $U(L)$ gauge transformations $W_{m}$ and $W_{n}$
(details in the Appendix C):

\bea  \label{gt}
&&|\phi_{m}^{a}(\vec{\lambda})\rangle \rightarrow|\phi_{m}^{b}(\vec{\lambda})\rangle (W_{m}(\vec{\lambda}))^{ba},
	\nonumber \\
&&|\phi_{n}^{a}(\vec{\lambda})\rangle \rightarrow |\phi_{n}^{b}(\vec{\lambda})\rangle (W_{n}(\vec{\lambda}))^{ba}.
\eea

Then the quantized character $\Theta$ can be defined between any
two eigenspaces associated with eigenvalues $E_\pm$.
$\Delta$ in Eq.~(\ref{theta}) is replaced by $\Delta_{\rm{D}}$, and
$\mathcal{F}$ is defined as $ \frac{1}{L}  {\rm{Tr}}\{\mathcal{F}_- -\mathcal{F}_+ \}$,
where $\mathcal{F}_{\pm} = \frac{1}{2}F_{\pm}^{\mu\nu} d\lambda^{\mu}\wedge d\lambda^{\nu}$
with $F^{\mu\nu}_{\pm} = \partial^\mu
\mathcal{A}^{\nu}_{\pm} - \partial^\nu \mathcal{A}^{\mu}_{\pm} - i \left[\mathcal{A}^\mu_{\pm}, \mathcal{A}^\nu_{\pm} \right]$
being the non-Abelian Berry curvatures.
$z(t) = \exp \{\frac{1}{L}{\rm Tr} \ln (UV^{\dag})\}\}$
defines a closed curve
in the complex plane, when $\vec{\lambda}$ completes a close loop.
Therefore, $W[z] = \int\frac{-i}{L} {\rm Tr} \{\ln UV^{\dag})\}$
is a winding number of $z$ relative to the origin of the complex plane, which
is quantized and play the counterpart of $\int_{\partial \mathcal{M}}d ~{\rm{arg}}
\langle\phi_+|\dot{\phi}_-\rangle$ in non-degenerate case.
Therefore Eq.~(\ref{theta}) still holds for degenerate system.

\paragraph*{Discussion and conclusions---}
Based on the gauge invariant quantum geometric potential, we
define a new quantized character $\Theta$ for both
non-degenerate and degenerate quantum systems.
It is a quantum analogy to the Gauss-Bonnet theorem for a manifold
with boundary.
This character is fundamentally different from the Chern number
which is quantized for the bundle based on a manifold without
boundary.
Furthermore, $\Theta$ is an inter-level index, while the Chern
number is an intra-band (level) property.
We speculate that this quantized inter-level character can be
further applied to the study of quantizations of physical
observables in topological physics and quantum adiabatic condition.

\paragraph*{Acknowledgement --- }
We thank X. L. Li, J. McGreevy, L. Ni, X. F. Zhou for helpful discussions.
C. X., J. W., and C.W. acknowledge the support from AFOSR FA9550-14-1-0168.
C. W. also acknowledges the support from the National
Natural Science Foundation of China (11729402).


\appendix
\begin{widetext}
\section{Time Evolving Equation for Degenerate system}
As discussed in the article, the solution to the time dependent Schr\"odinger equation can be expanded by $|\eta_{m}^a\rangle$ defined in Eq.~(\ref{eta}) in the main text as
\begin{equation}
	|\psi(t)\rangle =  c_{m}^a(t) |\eta_{m}^a(t)\rangle,
\end{equation}
or with $|\tilde{\phi}_{m}^a\rangle = |\phi_{m}^b(t)\rangle(\Omega_{m}(t))^{ba}$ as
\begin{equation}
	|\psi(t)\rangle =  c_{m}^a(t) \exp \{-i\int_0^t E_m(\tau) d\tau\}|\tilde{\phi}_{m}^a\rangle,
\end{equation}
where $\Omega_m$ is defined in Eq.~(\ref{omega}) in the main text. It can be shown that $\langle \tilde{\phi}_{m}^a| \dot{\tilde{\phi}}_{m}^a\rangle = 0$,
because
\bea
	\langle \tilde{\phi}_{m}^a| \dot{\tilde{\phi}}_{m}^a\rangle &=& (\Omega_{m}^\dag)^{ac}\langle \phi_{m}^c|\dot{\phi}_{m}^b\rangle (\Omega_{m})^{ba} +  (\Omega_{m}^\dag)^{ac}\langle \phi_{m}^c|\phi_{m}^a\rangle (\dot{\Omega}_{m})^{ba}  \nonumber
\\
&=& (\Omega_{m}^\dag)^{ac}(-i\mathcal{A}_{m})^{cb} (\Omega_{m})^{ba} + (\Omega_{m}^\dag)^{ac}\delta^{ca} (i\mathcal{A}_{m})^{ab}(\Omega_{m})^{ba} = 0.
\eea
Solving the time dependent Schr\"odinger equation $i\partial_t |\psi(t)\rangle = \hat{H}(t)|\psi(t)\rangle$:
\begin{equation}
	i\{\dot{c}^a_m|\tilde{\phi}_{m}^a\rangle -iE_m(t)c_m^a|\tilde{\phi}_{m}^a\rangle + c_m^a|\dot{\tilde{\phi}}_{m}^a\rangle\} \exp \{-i\int_0^t E_m(\tau) d\tau\} = E_m^a c_{m}^a \exp \{-i\int_0^t E_m(\tau) d\tau\}|\tilde{\phi}_{m}^a\rangle.
\end{equation}	
Left multiply $\langle \tilde{\phi}_m^a|$ to the equation above, one obtains
\begin{equation}
	i \{ \dot{c}_m^a - iE_m(t) \}\exp \{-i\int_0^t E_m(\tau) d\tau\} + \sum_{b,n,n\neq m}  i{c_n^b(t)}\langle\tilde{\phi}_m^a|\dot{\tilde{\phi}}_n^b\rangle\exp \{-i\int_0^t E_n(\tau) d\tau\} = E_m(t)c_m^a(t) \exp \{-i\int_0^t E_m(\tau) d\tau\}.
\end{equation}
Then one arrives at
\begin{equation}
	\dot{c}_{m}^{a}(t) = - \sum_{n,n\neq m}(\exp \{i\int_0^t\epsilon_{mn}(\tau)d\tau\}\langle\tilde{\phi}_m^a|\dot{\tilde{\phi}}_n^b\rangle) c_{n}^{b}(t).
\end{equation}
Therefore the time evolving equation of Eq.~(\ref{eq:time}) in the main text can be obtained,
\begin{equation}\label{time}
	\dot{c}_{m}^{a}(t) = - \sum_{n\neq m}\exp \{i\int_0^t\epsilon_{mn}(\tau)d\tau\}(\Omega_{m}^{\dag}T_{mn}\Omega_{n})^{ab}c_{n}^{b}(t),
\end{equation}
with $\epsilon_{mn}(\tau) = E_+(\tau) - E_-(\tau)$.
Therefore
\begin{equation}
\dot{c}_{m}^{a} = - \sum_{n\neq m}
\exp \left \{i\int_0^t(\epsilon_{mn}(\tau)+\Delta_{\rm{D},mn}(\tau)) d\tau
\right \}
\times (\bar{\Omega}_m^\dag \bar{U}_{mn}S_{mn}\bar{V}_{mn}^\dag
\bar{\Omega}_n)^{ab}c_{n}^{b}(t).
\end{equation}

With the gauge invariant $\Delta_{{\rm{D}}}$ in the degenerate systems
Eq.~(\ref{delta}), we can further revise the QAC for the quantum
degenerate systems.
For an adiabatic process, all the $c_m^b(t)$'s are nearly time-independent,
because $|\eta_m^b(t)\rangle$ are already the adiabatic evolution states.
If one further assumes that $\epsilon_{mn}(t)$,
$\Delta_{{\rm D},mn}(t)$, $S_{mn}$ and
$(\bar{\Omega}_m^\dag \bar{U}_{mn}S_{mn}\bar{V}_{mn}^\dag \bar{\Omega}_n)^{ab}(t)$
are slow varying variables, and the system is initially prepared
in the states $|\eta^a_{k}(0)\rangle$,
then the time-evolving part is approximately controlled by
$\exp \{i (\epsilon_{mn} + \Delta_{{\rm D},mn} )t \}$.
With these conditions, the QAC for the degenerate systems
can be expressed as $\forall m \neq n$
\be
	\frac{\max(S_{mn})}{|\epsilon_{mn} + \Delta_{{\rm D},mn}|} \ll 1,
\label{adc1}
\ee
where $\max(S_{mn})$ is the maximum value of the singular values of
the transition matrix $T_{mn}$.
Physically, the $\max(S_{mn})$ represents
the most possible channel in the process of transition.

To illustrate how the degenerate QAC Eq.~(\ref{adc1}) works, we construct
a two-level toy model as follows:
\begin{equation}
H(t) = \left[ \begin{matrix}
	\vec{n}_1(t)\cdot \vec{\sigma} & \\
	                          & \vec{n}_2(t)\cdot \vec{\sigma}
\end{matrix}\right].
\end{equation}
If $\vec{n}_1 = \vec{n}_2 = (\sin\theta\cos(\omega t), \sin\theta\sin(\omega t), \cos \theta)$,
then $H$ is simply a double copy of Rabi model.
$\Delta_{\rm{D}}$ can be calculated by Eq.~(\ref{delta}),
and the result is $(1-2\cos^2(\theta/2))\omega$.
After extracting the phase term $i\int_0^t \Delta_{\rm{D}}(\tau)d\tau$,
the remaining part $\bar{\Omega}_+^\dag \bar{U}S\bar{V}^\dag \bar{\Omega}_-$
is a constant, and $S$ is also a contant matrix $\sin(\theta)\omega/2\cdot \mathbb{I}_{2\times 2}$,
so that we can use Eq.~(\ref{adc1}) to judge
the adiabaticity as
\begin{equation}
	\frac{|\sin(\theta)\omega/2|}{|2+(1-2\cos^2(\theta/2))\omega|} \ll 1.
\label{adc3}
\end{equation}
When $\theta \to 0^+$, Eq.~(\ref{adc3}) breaks down if $\omega \simeq 2$, because the denominator goes to zero. This is expected since when $\omega$ matches the energy gap, the resonance happens so that the system is not adiabatic anymore.

Besides $\Delta_{\rm{D}}$, one can also define other gauge invariants within the general time-dependent problem described above.
Every single element of the matrix, $(\Omega_{m}^{\dag}T_{mn}\Omega_{n})^{ab}$ can be evaluated as $\langle\tilde{\phi}_m^a|\dot{\tilde{\phi}}_n^b\rangle$, and it is also gauge invariant as long as the initial basis are fixed.
Similar as what we do in the non-degenerate case, we can separate the phase factor from $\langle\tilde{\phi}_m^a|\dot{\tilde{\phi}}_n^b\rangle$ as
\begin{equation}
	\langle\tilde{\phi}_m^a|\dot{\tilde{\phi}}_n^b\rangle = \exp \{i\int_0^t \Delta^{ab}_{mn} d\tau \} |\langle\tilde{\phi}_m^a|\dot{\tilde{\phi}}_n^b\rangle|
\end{equation}
with $\Delta^{ab}_{mn} = \frac{d}{dt} \arg(\langle\tilde{\phi}_m^a|\dot{\tilde{\phi}}_n^b\rangle)$.
Then Eq.~(\ref{eq:time}) can be rewritten by using $\Delta^{ab}$ as
\begin{equation}
	\dot{c}_m^a(t) = -\sum_{m\neq n} \exp \{i\int_0^t (\epsilon_{mn}(\tau) + \Delta^{ab}_{mn}(\tau)) d\tau \}|\langle\tilde{\phi}_m^a|\dot{\tilde{\phi}}_n^b\rangle| c_n^b(t).
\end{equation}
If one further assume $\epsilon_{mn}(t) = \epsilon_{mn}$, $|\langle\tilde{\phi}_m^a|\dot{\tilde{\phi}}_n^b\rangle|$ and $\Delta^{ab}_{mn}$ are slow varying varibles, the adiabatic condition can be deduced as
\begin{equation}\label{adc2}
	\frac{|\langle\tilde{\phi}_m^a|\dot{\tilde{\phi}}_n^b\rangle|}{|\epsilon_{mn} + \Delta^{ab}_{mn}|} \ll 1 ~~~\forall a, b,m\neq n.
\end{equation}
$|\tilde{\phi}_m^a\rangle$ are adiabatically evolved basis, so that the meaning of QAC Eq.~(\ref{adc2}) is that all the transitions between any two adiabatically evolved basis with different energies are all very weak, so that this degenerate system can evolve adiabatically.

\section{Ambiguity of  the Singular Value Decomposition (SVD)}
For a general $l\times l$ matrix $C$, when applying SVD to it, one will obtain $(C)^{ab} = (U)^{ad}(\Lambda)^{d}(V^{\dag})^{db}$, with $l$ non-negative singular values $\Lambda_{d}$ (a, b and d vary from $1\rightarrow l$) and $U$ and $V$ being unitary matrices. SVD has its intrinsic ambiguity that comes from the unitary matrices $U$ and $V$. In the case that all the singular values are positive, one can insert two diagonal matrices as:
\begin{equation}
	(C)^{ab} = (U)^{ad}(\Lambda)^{d}(V^{\dag})^{db} = (U)^{ad}e^{i\lambda_d}(\Lambda)^{d}e^{-i\lambda_d}(V^{\dag})^{db}
\end{equation}
with $\lambda_d$ being any real numbers.
After the insertion, one can define $(U^{\prime})^{ad} = (U)^{ad}e^{i\lambda_d}$ and $(V^{\prime})^{ad} = (V)^{ad}e^{i\lambda_d}$, so that $C = U^{\prime}\Lambda {V^{\prime}}^{\dag}$ which is also a valid SVD of $C$. Therefore SVD has its intrinsic ambiguity of the choice of the unitary matrices, but there is neither ambiguity of the singular values nor ambiguity of the product of $U$ and $V^{\dag}$ in this case.

When the singular values of a matrix $C$ contain a zero or multiple zeros, there are further ambiguities. For example, if $C$ is decomposed as $C = U\Lambda V^{\dag}$ and the $n^{th}$ singular value is zero, then one can also insert two diagonal matrices as
\begin{equation}
	(C)^{ab} = (U)^{ad}(\Lambda)^{d}(V^{\dag})^{db} = (U)^{ad}e^{i\lambda_d}(\Lambda)^{d}e^{-i\lambda_d^{\prime}}(V^{\dag})^{db}
\end{equation}
with $\lambda_d$ and $\lambda_d^{\prime}$ being any real numbers and $\lambda_d = \lambda_d^{\prime}$ if $d\neq n$. Because the $n^{th}$ singular value is zero, $\lambda_n$ and $\lambda_n^{\prime}$ do not have to be equal. Define $(U^{\prime})^{ad} = (U)^{ad}e^{i\lambda_d}$ and $(V^{\prime})^{ad} = (V)^{ad}e^{i\lambda_d^{\prime}}$, so that $C = U^{\prime}\Lambda {V^{\prime}}^{\dag}$, however $UV^{\dag} \neq U^{\prime}{V^{\prime}}^{\dag}$.

\section{ Proof of the Gauge Invariance of the Quantum Geometric Potential $\Delta_{\rm{D}}$}

As mentioned in the article, $\Delta_{\rm{D}}$ is gauge invariant under any independent $U(L)$ gauge transformations $W_{m}$
\begin{equation}
	|\phi_{m}^{a}(\vec{\lambda})\rangle \rightarrow |\phi_{m}^{b}(\vec{\lambda})\rangle (W_{m}(\vec{\lambda}))^{ba}.
\end{equation}
Under the gauge transformations above, $\mathcal{A}_{m}$, $T_{mn}$ and $U_{mn}V_{mn}^{\dag}$ transform as follows:
\begin{eqnarray}
	\mathcal{A}_{m}^{\mu} &\rightarrow & W_{m}^{\dag}\mathcal{A}_{m}^{\mu}W_{m} + i W_{m}^{\dag} \partial_{\lambda^{\mu}}W_{m}\label{gaugetrans1}\\
	T_{mn} &\rightarrow & W_{m}^{\dag} T_{mn} W_{n}\label{gaugetrans2} \\
	U_{mn}V_{mn}^{\dag} &\rightarrow & W_{m}^{\dag}U_{mn}V_{mn}^{\dag}W_{n}.
\label{gaugetrans3}	
\end{eqnarray}
($U_{mn}$ and $V_{mn}^{\dag}$ are the unitary matrices come from the SVD of $T_{mn}$; $\mathcal{A}_{m}$ and $T_{mn}$ are introduced in Eq.~(\ref{A}) and Eq.~(\ref{tran}) in the main text). $\Delta_{\rm{D}}$ is carried out as
\begin{equation}
	\Delta_{{\rm{D}}, mn} = \frac{1}{L} {\rm{Tr}} \{(\mathcal{A}_{n}-\mathcal{A}_{m})+i\frac{d}{dt}(-\ln(U_{mn}V_{mn}^\dag))\},
\end{equation}
and under the gauge transformations $W_{m}$
\begin{equation}\label{gtod}
\Delta_{{\rm{D}},mn}\xrightarrow{W_{m}}{} 	\frac{1}{L} {\rm{Tr}} \{(W_{n}^{\dag}\mathcal{A}_{n}W_{n} + i W_{n}^{\dag}\dot{W}_n -W_{m}^{\dag}\mathcal{A}_{m}W_{m} - i W_{m}^{\dag}\dot{W}_m )+i\frac{d}{dt}(-\ln(U_{mn}V_{mn}^\dag) -\ln(W_m^\dag) - \ln(W_n)) \}.
\end{equation}
We have used the fact that ${\rm{Tr}} \{ \ln(AB)\} = {\rm{Tr}}\{\ln(A) \} + {\rm{Tr}}\{\ln(B)\}$ if $A, B \in U(L) $ in the equation above.
$W_{m}^{\dag}\mathcal{A}_{m}W_{m}$ are similarity transformations, so that the trace remains the same as before. If $A \in U(L)$, then ${\rm{Tr}} \{\ln(A)\} = \ln(\det(A))$, so that $\frac{d}{dt} {\rm{Tr}} \{\ln(A)\} = \frac{d}{dt} {\rm{Tr}} \{\ln(\Lambda)\}$ with $V\Lambda V^\dag = A$ and $\Lambda$ being diagonal. If $V\Lambda V^\dag = A \in U(L)$, then
\begin{equation}
	{\rm{Tr}} \{A^\dag \dot{A}\} = {\rm{Tr}} \{V\Lambda^\dag (V^\dag \dot{V})\Lambda V^\dag + V\dot{V}^\dag + V \Lambda^\dag \dot{\Lambda}V^\dag\} = {\rm{Tr}} \{\Lambda^\dag \dot{\Lambda} \} = {\rm{Tr}} \{\Lambda^{-1} \dot{\Lambda} \} = \frac{d}{dt} {\rm{Tr}} \{\ln(\Lambda)\},
\end{equation}
so that if $A \in U(L)$, $\frac{d}{dt} {\rm{Tr}} \{\ln(A)\} = {\rm{Tr}} \{A^\dag \dot{A}\}$. Then Eq.~(\ref{gtod}) can be simplified as
\begin{equation}
	\Delta_{\rm{D}} \xrightarrow{W_{m}}{}  \Delta_{\rm{D}} + \frac{1}{L} {\rm{Tr}} \{-i\frac{d}{dt}(\ln(W_n^\dag) + \ln(W_n) + \ln(W_m^\dag) - \ln(W_m^\dag)) \} = \Delta_{\rm{D}}.
\end{equation}
Therefore $\Delta_{\rm{D}}$ is gauge invariant under a $U(L)\times U(L)$ gauge transformation.

As for the case that the degeneracies of these two eigenspaces
are different, one can still define the gauge invariant as
\begin{equation}
	\Delta_{{\rm{D}},mn} = -\frac{i}{{\rm{min}}(L_m, L_n)} {\rm{Tr}}\{\dot{X}_{mn}X_{mn}^{\dag}\},
\end{equation}
where $X$ is $\Omega^{\dag}_m U_{mn}V^{\dag}_{mn}\Omega_n$,
and $L_m$ and $L_n$ are the degeneracies of these two eigenspaces (suppose $L_m < \L_n$).
As mentioned in the main text of this article, $V^\dag V = \mathbb{I}_{L_m\times L_m}$ is an
identity matrix, while $V V^\dag$ is not.
$\mathcal{A}_{m}$, $T_{mn}$ and $U_{mn}V_{mn}^{\dag}$
transform as same as Eq.~(\ref{gaugetrans1}), Eq.~(\ref{gaugetrans2}) and Eq.~(\ref{gaugetrans3}),
so that under the gauge transformation:
\begin{eqnarray}
	\Delta_{{\rm{D}},mn}  &=& -\frac{i}{L_m} {\rm{Tr}} \{-i\mathcal{A}_m + \frac{d}{dt}(U_{mn}V_{mn}^{\dag})(V_{mn}U_{mn}^\dag) +i U_{mn}V_{mn}^\dag \mathcal{A}_n V_{mn}U_{mn}^{\dag}  \}\\
	&\xrightarrow{W_m}{}& -\frac{i}{L_m} {\rm{Tr}} \{ -i W_m^{\dag}\mathcal{A}_m W_m + W_m^{\dag}\dot{W}_m + \dot{W}_m^{\dag}W_m + \frac{d}{dt}(U_{mn}V_{mn}^{\dag})(V_{mn}U_{mn}^\dag) \\
	& &+  U_{mn}V_{mn}^{\dag} \dot{W_n}W_n^{\dag} V_{mn}U_{mn}^\dag  + W_m^{\dag}U_{mn}V_{mn}^{\dag}W_n(iW_n^{\dag}\mathcal{A}_n W_n - W_n^{\dag}\dot{W}_n)W_n^{\dag} V_{mn}U_{mn}^{\dag}W_m\}\\
	&=& -\frac{i}{L_m} {\rm{Tr}}\{ -i W_m^{\dag}\mathcal{A}_m W_m +  \frac{d}{dt}(U_{mn}V_{mn}^{\dag})(V_{mn}U_{mn}^\dag) +i W_m^{\dag}U_{mn}V_{mn}^{\dag} \mathcal{A}_n V_{mn}U_{mn}^{\dag} W_m \}\\
	&=& \Delta_{{\rm{D}},mn}.
\end{eqnarray}
Therefore the gauge invariance is verified.
Note that some terms in the equations above, like $U_{mn}V_{mn}^\dag \mathcal{A}_n V_{mn}U_{mn}^{\dag} $ and $U_{mn}V_{mn}^{\dag} \dot{W_n}W_n^{\dag} V_{mn}U_{mn}^\dag$ are in fact not similarity transformations of $\mathcal{A}_n$ and $\dot{W_n}W_n^{\dag}$.

\end{widetext}

\end{document}